# Generation of pedagogical content based on the learning style of learners in a dynamic adaptive hypermedia environment

A.BOUCHBOUA, R.OUREMCHI
LTTI, Department of Electrical and Computer Engineering
ESTF, Sidi Mohamed Ben Abdellah University
Fez, Morocco
ahmed.bouchboua@usmba.ac.ma,
rabah.ouremchi@usmba.ac.ma

F.MESSAOUDI, M.EL GHAZI
LTTI, Department of Electrical and Computer Engineering
ESTF, Sidi Mohamed Ben Abdellah University
Fez, Morocco
faycal.messaoudi@usmba.ac.ma,
elghazimo@hotmail.fr

*Abstract*—several researches in psychology and science of education affirm the impact of learning style on the learning process and encourage its taking into account in the teaching strategies in order to facilitate the task for learners and improve their results.

This article deals with the relationship between the characteristics of the learner, the teaching materials and the context in which takes place the learning in order to allow a better adaptivity. The latter is ensured thanks to the most important element of our system, which is the generator of course, the latter allows you to offer a hypermedia virtual, therefore the pages and the links will be built dynamically, taking into account the learning style and the cognitive status of the learner. As well and after having completed a first filter on the fragments, in order to select those corresponding to the course, there will be applied a second filter to select the fragments corresponding to the learning style of the learner to retain only those in accordance with responsive to the level of knowledge required.

## I. Introduction

The hypermedias have opened a new line of research in the field of education systems assisted by computer. In this framework, they have emerged three types of systems: first the hypermedia so-called conventional, then the adaptive hypermedia and finally the hypermedia adaptive dynamic.

The work in this area are to design of computer systems based on models of learners whose purpose is to assist the latter in the context of a know abundant and widely distributed. These learners are different in terms of training, attitudes, skills, motivations, preferences, expectations and needs.

Our research works are located in this path, in order to take into account the real needs of learners for a better adaptation of content and learning. We then we are interested in the question of access to knowledge in a process of personalized learning, individualized and adapted to the needs of learners.

In the first part of this article, we will introduce the models of learning style and we will celebrate the choice of the model that we have adopted. In the second part, we describe the architecture of our system adaptive hypermedia dynamic. The second part will focus on the mechanisms of adaptation of content to the educational profile of learners. We will finish our work by a conclusion.

## II. Models of learning style

Each individual has a personal style of reading and learning. A manner which to him is own to organize the concepts and information. This is what is known in pedagogy and psychology in terms of: learning styles. This justifies that a situation of teaching cannot be perceived in the same way by all learners.

Several models of learning style have emerged which are grouped into three typologies:
- The models of learning style who are interested in the preferences for the conditions of teaching and learning.
- The models of learning style who are interested in the way in which the learner processes the information, in terms of average privileged.
- The models of learning style which treaty the personality of the learner.

The hypermedia systems dynamic adaptive must identify - in a prime time - for each learner: its learning objectives [1], its level of knowledge, its preferences [2], its stereotypes, its choice cognitive and its learning styles [3].

The levy of these characteristics is done, either, at the beginning of the session of learning, either, by the observation of the behavior of the learner in the course of the session, that is, at the end of the session of learning. The method of transformation of observations in educational information exploitable is named profiling.

We are interested, in this part, to the study of the profiling process of learners using the measure of learning style. This measure is based on the index of learning styles"

Index of Learning Styles (ILS) "established by Felder and Silverman [4]. They considered that the approaches to collect and treat mentally information differs from one person to another. These approaches can be classified according to four dimensions.

## III. EASE OF USE PRELUDE OF THE QUESTIONNAIRE OF FELDER-SILVERMAN

The questionnaire of Felder-Silverman that we have translated into French contains 44 questions. For each question, the learner must choose a response between a and b. The 44 questions are divided into four groups of 11 questions each.

Each group of questions defines a dimension for the cognitive model of the learner who is therefore composed of 4 dimensions according to Felder-Silverman:

### A. Dimension 1: Processing of information

The first dimension is the dimension of the reflection and the processing of information by the learner. It varies from the thought to the asset. The active learners succeed better by engagement in an activity (collective or individual) or by a discussion of the concept taught. A learning system it should, therefore, give more interest to the practical aspect of the teaching and the implementation of co-operative activities and collaborative.

The thoughtful learners prefer learning by introspection (observe, listen to, etc. ). To be more effective with a learner thought, the educational device should be, most of the time, based on the theory, the definitions and demonstrations.

### B. Dimension 2: Reasoning

This second dimension represents the reasoning. It varies from deductive to inductive. The syntactics learners prefer from the principles in order to deduce the consequences or applications. The learners inductive, by contrast, prefer to go for facts and examples to identify the principles. An educational device adapted to learners syntactics should begin by definitions and theories and advancing toward the practice.

### C. Dimension 3: Perception of information

The third dimension is to represent the manner with which the learner prefers collecting the information. This is the sensory dimension. It varies from the visual to the verbal. A visual learner prefers a teaching, using images, charts, graphs and animations. By contrast, a learner record preferred a teaching using the texts, words, readings and discussions.

### D. Dimension 4: Progress toward understanding

This dimension defines the way that the learner prefers to advance the learning of a course. It varies between global and sequential. A learner prefers sequential progress through successive stages. By contrast, a global learner prefers to choose freely his journey to make large jumps in function of the context.

## IV. MEASUREMENT OF A DIMENSION

To assign a dimension to a learner, using the questionnaire of Felder-Silverman , it is sufficient to count the number of answers "a" and the number of answers "b" on the 11 questions corresponding to the dimension as it is shown on Table 3 and calculate the difference between these two numbers. Are M and N respectively these two numbers. The difference M-N allows to situate the learning style of the learner. A negative number indicates that the learner is close to the end b and vice versa.

This measurement is between 11 (all the answers of the learner are equal to a) and -11 (all responses are equal to b). The learner may be close to the end (b) if he has obtained a negative number and vice versa.

You can assign to this extent a degree of confidence calculated by $Cf = |M-N|$ according to table 4, below:

TABLE I. DEGREE OF CONFIDENCE FOR THE MEASUREMENT OF A DIMENSION OF LEARNING STYLE

| Degree of Confidence | Meaning |
|---|---|
| 1 - 2 - 3 | Uncertain |
| 4 - 5 - 6 - 7 - 8 | Moderate |
| 9 - 10 - 11 | Fort |

## V. IMPLEMENTATION OF THE STUDY

After the setting on line of the questionnaire of Felder-Silverman, intended for the students of the higher school of technology of Fez, the results have been exported by the administration interface of LIMESURVEY. The analysis of these results has allowed us to show the rate of reliability of the measuring instrument used in the experiment. For this purpose, we calculated the coefficient of Cronbach's, to assess the reliability of the measures offered by the questionnaire. Indeed, in [5] the authors state that the acceptance threshold of the coefficient of Cronbach's depends on the type of questionnaire. In the case of a questionnaire which measures a level of knowledge, the minimum threshold of acceptance is fixed at 0.75 and in the case of a measure of preferences or of attitude to 0.50.

On the SPSS software, our survey has obtained a α coefficient of Cronbach's equal to 0.7609 and higher to 0.50 , which indicates a reliability rate of decent for a measure of the learning style.

The α coefficient Cronbach's is defined as follows:

$$\alpha = \frac{k}{k-1}\left(1 - \frac{\sum_{i=1}^{k}\sigma_{Y_i}^2}{\sigma_X^2}\right)$$

Where k is the number of items, is the variance of the total score and α is the variance of the item i.

## VI. STATISTICAL ANALYSIS OF THE RESULTS OBTAINED IN OUR STUDY

The primary objective of this study is to measure the learning style of each learner and then, to reveal the profile the more popular, which will be assigned to all new registered on our system, who have not gone through the questionnaire of Felder.

After the analysis of our results, we found that the target population was made up of multiple and different learning styles.

According to the proposed questionnaire, the learner must complete the 44 questions, which each group of 11 questions is represented by a dimension and in each dimension we find two different values. We can deduce that $4^2 = 16$ Learning Styles as possible in our study.

The study we conducted, on each dimension to share, does not really reflect the learner profiles. Then, the results obtained up to now, remain inadequate. To know the style of a learner, it must assign a value for each dimension, that is to say, a profile is represented by a combination of four different values.

For this purpose, we have observed the four dimensions for each learner. The results shown in the figures below, represent the strength of learners in each learning style.

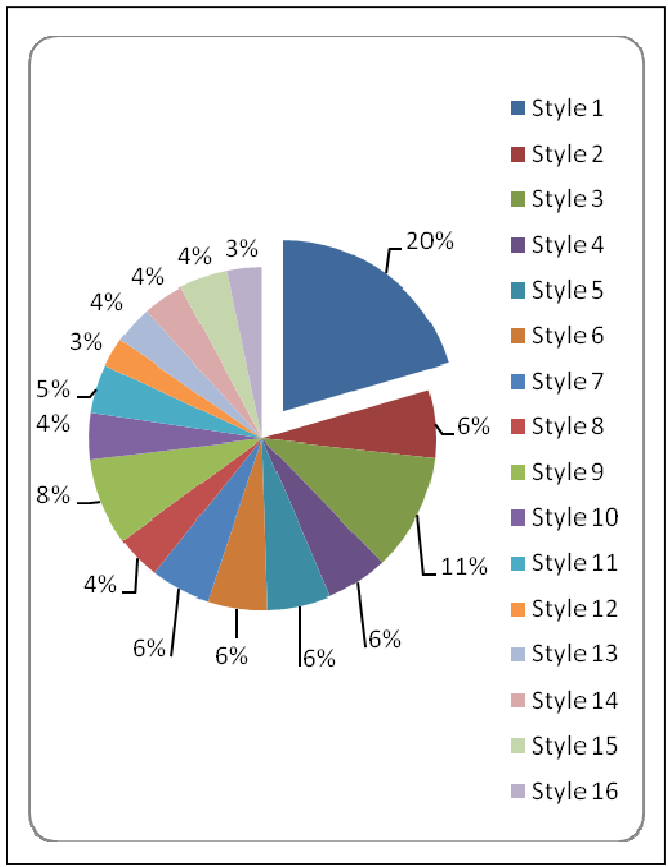

Figure 1. Distribution of learners in each learning style.

We note, according to the figure above, that the style 1 is the most popular among the learners, then the style 3 and then the style 9, etc. etc. More than 20% of the population belong to this style. This result present the objective of our study, as we have mentioned previously, this style will be assigned by default to the new learners who do not wish go the questionnaire of Felder-Silverman. The style 1 is composed of four dimensions:
- The sensory dimension: Verbal
- The dimension Progress: Sequential
- The dimension Reflection: Active
- The dimension Reasoning: Inductive

We can deduce that the learners who belong to the style 1 prefer educational resources textual and audible, to navigate and move step by step in a sequence of learning. They also prefer practical activities, individual or collective and begin their sequences by examples, of the facts; then, practices and then the theory.

VII. ARCHITECTURE OF OUR SYSTEM

The architecture of our adaptive system dynamic is composed essentially of a domain model, a pedagogical model, a model of the learner, a multimedia database and a generator of course. [6]

The first: still called the model of knowledge. The objective is to determine the relevant concepts and their relationships and to provide a comprehensive structure of the domain of learning. This model is focused on the design of a creation environment designed to authors for the production of educational content dedicated to learners. [7]

The second model that we have invented and that is the educational model. He detailed how to model the teaching strategies used by teachers, during the presentation of the teaching content, to learners. It also stores the parameters used to introduce the cognitive status of the learner. It focuses on a unit of prediction to predict the following concept which could be visited by the learners and allows also to promote the autonomy of the learner who has a need to have a feed-back, in order to allow him to estimate of how effective the effort he provides and those that remain to provide. [8]

The third: is the model of the learner which allows to take into account the different characteristics of the learner, namely, its personal information, its needs, preferences, his attitudes and skills. [9]

In our approach, the learner needs to be modeled, in the first place, by characteristics which are grouped in a facet of identification with the personal information such as: the login, password, the name, the first name, age, e-mail, etc... And in the second place, by the purpose and need of the learner to the knowledge that may be general or depth. The acquisition of such information can only be done with a questionnaire that the learner must fulfill during his first contact with the system. Then, a model of knowledge which is defined by the score obtained for each concept or element of knowledge. The modeling of this model is done in different ways in which we have chosen the method expertise-partielle (Overlay) or the state of knowledge of the learner is represented as a sub-set of knowledge from the domain model.

In effect, this model provides a bridge between the learner and the different concepts that he has studied. This relationship has a numeric value summarizing the rate of assimilation of pre-requisite needed to address this concept. Finally, a cognitive model to detect the profile or the learning style of the learner.

The fourth component: is the multimedia database. This database contains the concepts and the documents to be presented to the learner, in function of the domain model [10]. These documents are characterized by attributes to identify those who should be presented to the learner.

The fifth and the last component: known as the generator of course, is regarded as one of the most important parts in our system, because it connects the different parts of the system defined previously and allows to offer a virtual hypermedia.

Figure 2 presents the overall architecture of our system adaptive hypermedia dynamic.

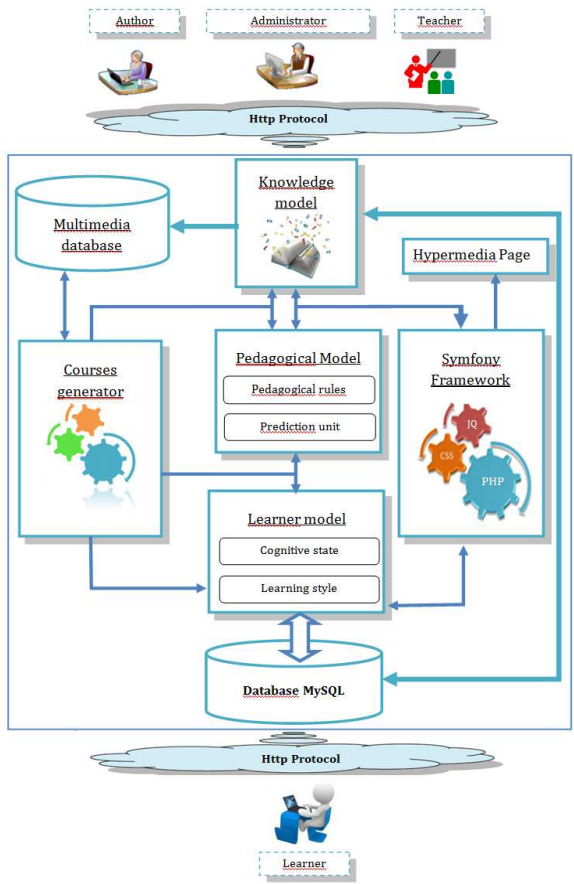

Figure 2. The architecture of our system adaptive hypermedia dynamic

## VIII. ADAPTATION MECHANISM ACCORDING TO THE LEARNING STYLE OF THE LEARNER

We can summarize the mechanisms of adaptation of the pedagogical content according to the learning style of the learner in six steps:

Step 1: The system begins by identifying the learner, if it is of its first use, it is responsible for collecting and save all the personal information of the learner including its id and its password on the one hand and on the other hand, the information related to its educational enrollment.

Step 2: The learner is found before two separate choice either: answer the questionnaire in Felder-Silverman and go on to the fourth step; this questionnaire is a set of questions of psychological aspect whose purpose is to detect easily the learning style, preferences, and attitudes of the learner. Otherwise, the generator of course passes to the third stage.

Step 3: In the case where the learner exceeds the questionnaire, the system must assign it a style of learning by default that we have defined, previously, and go to the fifth step.

Step 4: The set of answers of the learner to the questionnaire will allow our system to measure and define the degrees of preference for each dimension of the Felder-Silverman model.

Step 5: This step allows you to measure the parameters of adaptation of content according to the degrees of preference.

Step 6: When the learner specifies its objective and indicates to the system the course that he wants to follow, the generator of the course will recover:

- The concept to learn,
- The level of knowledge of the learner on this concept,
- The structure of the page and the hypertext links own the concept chosen, depending on the results of the measurement of parameters of adaptation

Figure 3, summarizes the adaptation mechanism according to the style of learning of the learners:

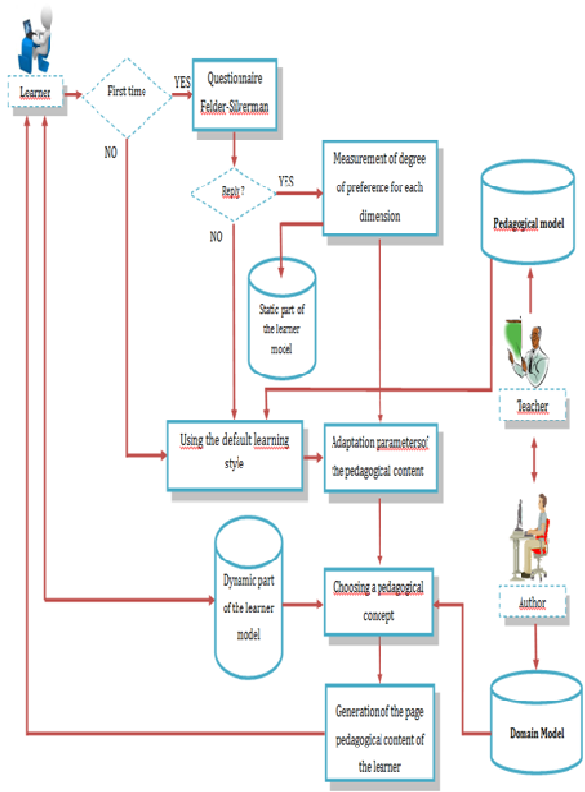

Figure 3. Mechanisms of adaptability of the content according to the style of learning

## IX. ADAPTATION MECHANISM ACCORDING TO THE COGNITIVE STATUS OF THE LEARNER

Our system allows the monitoring, supervision of the learner and the registration of all its traces, during its navigation on the hypermedia and during the procurement of pre-tests and post-tests relating to each learning objective. Therefore, the mechanism of adaptation of content, depending on the cognitive status, can be summarized in four steps:

Step 1: When a learner, already registered, chooses a course for the first time, the system is responsible to issue him a questionnaire; but this time, type of knowledge. The result of this questionnaire will enable the system to initialize the sub-model knowledge of the learner by assigning a level for him (Beginner, Intermediate, or Expert), if the learner exceeds this pre-test, the system will present the structure of the content by default.

Step 2: This step allows you to generate the structure of the teaching content, this generation is concentrated on three points:
- The relationship exists between the objectives and concepts stored on the domain model.
- The new cognitive status of the learner for each learning unit.
- The pedagogical rules stored on the pedagogical model.

Step 3: The system always gives the learner with the opportunity to to days his cognitive status in passing a post-test for each concept studied and after each session of learning.

Step 4: The system generates a new structure of the course to the learner, according to its last cognitive status, based on the domain model and the interaction of the learner with the system.

Figure 4, below, contains the steps that the generator of course can run to adapt the teaching content, depending on the cognitive status of the learner:

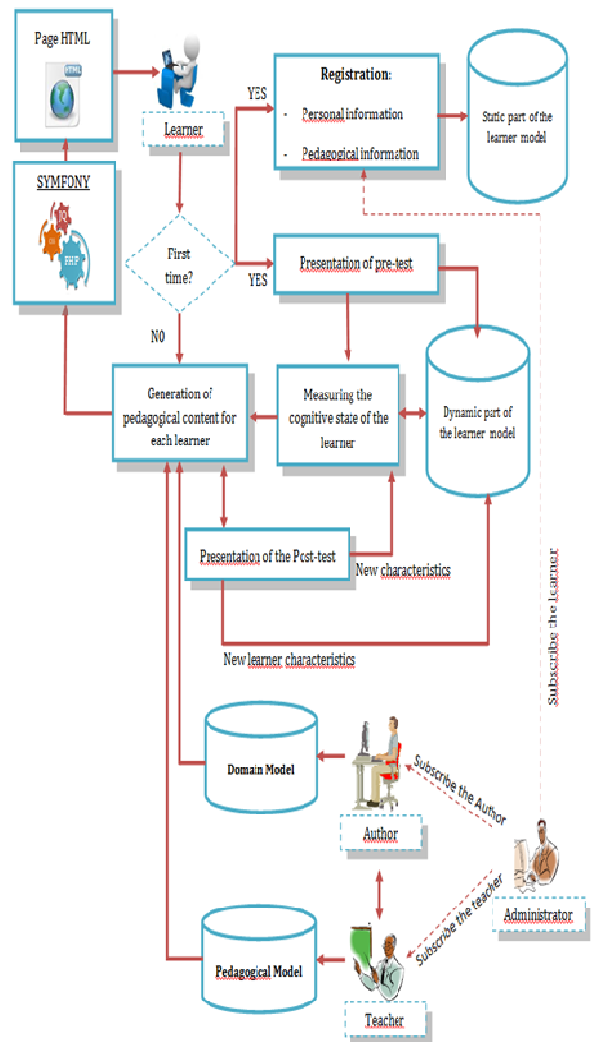

Figure 4. Mechanisms of adaptability of the content according to the cognitive status

## X. CONCLUSION

A hypermedia system dynamic adaptive is distinguished from a computer system classic by its architecture comprising a user template (model learner), a domain model and a pedagogical model. In this article, we introduced our system adaptive hypermedia dynamic based on the THEY (measurement of the learning style), which is able to determine the preferences of the learner, his desires, and his habits, and it has submitted its comprehensive architecture which is based on the overall architecture of adaptive hypermedia dynamic by improving the techniques of acquisition of a share and the characteristics of the model of the domain and of the model of the learner in order to obtain an adaptation at the level of the substance, and to better understand what the learner has understood, and on the other hand to adapt the contents of documents to its knowledge. We have also granted a importance to the mechanism of adaptation of courses according to the style of learning and the cognitive status of learners.